# First performance test of 30 mm iron-based superconductor single pancake coil under 24 T background field

Dongliang Wang[1,2,#], Zhan Zhang[3,#], Xianping Zhang[1,2], Donghui Jiang[4], Chiheng Dong[1], He Huang[1,2], Wenge Chen[4], Qingjin Xu[3,*], and Yanwei Ma[1,2,*]

[1] Key Laboratory of Applied Superconductivity, Institute of Electrical Engineering, Chinese Academy of Sciences, Beijing 100190, People's Republic of China

[2] University of Chinese Academy of Sciences, Beijing 100049, People's Republic of China

[3] Institute of High Energy Physics, Chinese Academy of Sciences, Beijing 100049, People's Republic of China

[4] High Magnetic Field Laboratory, Chinese Academy of Sciences, Hefei 230031, People's Republic of China

**Abstract**

The iron-based superconductor (IBS) single pancake coil (SPC) with 30 mm inner diameter was firstly fabricated and tested under 24 T background field. This SPC was successfully made using the 7-filamentary $Ba_{1-x}K_xFe_2As_2$ (Ba122) tape by wind-and-react method. This IBS coil show the highest $I_c$ value at magnetic field reported so far. For example, the transport critical current of this Ba122 SPC achieved 35 A at 4.2 K and 10 T, which is about half of that of short sample. This indicates that the non-insulation winding process together with the stainless-steel tape is suitable to the iron-based superconductor. Even more encouraging is the fact that the $I_c$ of this SPC is still as high as 26 A under 24 T background field, which is still about 40% of that at zero external magnetic field. These results clearly demonstrate that the iron-based superconductors are very promising for high-field magnet applications.

[#] Both authors contributed equally to this work.
[*] Author to whom correspondence should be addressed: E-mail: xuqj@ihep.ac.cn or ywma@mail.iee.ac.cn

## 1. Introduction

Since 2008, iron-based superconductors (IBS) [1] have attracted wide interest for both basic research and practical applications. Because they have high superconducting transition temperature ($T_c$), large critical current density ($J_c$) over 1 MA cm$^{-2}$ in thin films, very high upper critical field above 100 T and low anisotropy [2, 3]. After ten year's development, remarkable progress of high-performance Sr/BaKFeAs (Sr/Ba122) iron-based superconducting tape has been obtained [4-7]. Currently, record high transport $J_c$ value of Ba$_{1-x}$K$_x$Fe$_2$As$_2$ (Ba122) tapes is as high as 0.15 MA/cm$^2$ at 4.2 K and 10 T by hot-pressing process [8]. It is noteworthy that this hot-pressed sample also retained a high value of 5.4 ×10$^4$ A/cm$^2$ at 4.2 K and 27 T. This suggests that IBS is quite attractive for the construction of high-field magnets needed for nuclear magnetic resonance (NMR) spectrometers, particle accelerators, fusion reactors and also magnetic resonance imaging (MRI) systems.

For the practical application, the 100 m class Sr$_{1-x}$K$_x$Fe$_2$As$_2$ (Sr122) tapes have been fabricated in 2017 [9]. An average $J_c$ of 1.3 × 10$^4$ A/cm$^2$ at 4.2 K and 10 T was reached over the 115-m length, showing good longitudinal uniformity. Besides high current carrying capacity [10, 11], the Sr/Ba122 IBS tapes also showed good mechanical strength [11, 12], excellent reversible compressive strain (> 0.6%) [13, 14], and small anisotropy [8, 15]. However, good results of IBS coils were not yet achieved at both high and low magnetic field, even when the inner diameter of double pancake coil is as large as 71.5 mm [9].

In this paper, we successfully fabricated high transport $I_c$ 7-filamentary

Ba122/Ag/AgMn tape and single pancake coil (SPC). Though the inner diameter of Ba122 SPC is as small as 30 mm, its transport $I_c$ value still maintained about half of the short sample at 4.2 K and 10 T. Meanwhile, the $I_c$ value was still as high as 26 A under 24 T background field, which is about 40% of that at zero external magnetic field. Our results clearly indicate that the iron-based superconductors are very promising for high-field magnet applications.

## 2. 7-filamentary Ba122/Ag/AgMn tapes

*2.1. Experimental details*

The starting materials for preparing Ba122 precursor are small Ba fillings (99%), K pieces (99.95%), As (99.95%) and Fe (99.99%) powders [8]. To compensate the loss of K element during the sintering process, additional K around 20 at.% was added. The prepared powders were put into the milling jar and mixed by ball milling process. All procedures are handled in Ar atmosphere. After the ball milling, the powders were loaded into Nb tube, and then heat-treated at 900 ℃ for 35 h. The sintered precursor was ground into fine powder. To improve grain connectivity of the Ba122, 5 wt.% Sn powder was added and mixed with precursor powders in an agate mortar. The final powder was packed into silver tube, which was drawn into mono-filamentary wire. The mono-filamentary wire was then cut and bundled into AgMn0.4 tube and finally deformed into 7-filamentary Ba122/Ag/AgMn wire with diameter of 1.65 mm. This wire was then rolled into tape with thickness of 0.33 mm. Short 7-filamentary Ba122/Ag/AgMn samples were cut from the long tape, and then sintered at 880 ℃ for 0.5 h. The cross-section of the wire was observed using optical microscope. The temperature dependence of the resistivity was carried out using a four-probe method by PPMS (Model: PPMS-9). Transport critical current $I_c$ at 4.2 K and its magnetic field dependence were measured by the standard four-probe method with a criterion of 1 $\mu V$/cm. The magnetic field dependence of transport $I_c$ values for all samples were evaluated at the High Field Laboratory for Superconducting Materials (HFLSM) in Sendai.

*2.2. Results and discussions*

The Specification of $Ba_{0.6}K_{0.4}Fe_2As_2$/Ag/AgMn tape are shown in table 1. As shown from table 1, the ratio of non-superconductor to Ba122 superconductor is around 5. The transport $I_c$ value of 7-filamentary Ba122/Ag/AgMn tape is 72 A at 4.2 K and 10 T. The magnetic field dependence of transport $J_c$ data at 4.2 K for 7-filamentary Ba122/Ag/AgMn tape are shown in figure 1. The transport $J_c$ is about $3.01 \times 10^4$ A/cm$^2$ at 4.2 K and 10 T. This value is almost three times that of our previous 7-filamentary $Sr_{0.6}K_{0.4}Fe_2As_2$ tape [9]. The inset of figure 1 shows the optical micrograph on transverse cross section of sintered 7-filamentary Ba122/Ag/AgMn tape. It can be seen that the central Ba122 filament is thinner and broken, compared with the other 6 surrounded filaments. Obviously, this broken central Ba122 filament is harmful to the transport $I_c$ values and their uniformity. These problems should be further solved by optimizing the deforming process.

## 3. Ba122 single pancake coils

*3.1. Experimental details*

Ba122 single pancake coils (SPC) were designed by the Institute of High Energy

Physics (IHEP), Chinese Academy of Sciences. The SPCs were wound using non-insulation Ba122/Ag/AgMn tapes together with 0.1 mm stainless steel tapes by the wind-and-react method. The specifications of SPC are listed in table 2. The dimensions of SPC were 30 mm in inner diameter and 34.8 mm in outer diameter. Figure 2 (a) shows the outer view of Ba122 single pancake coil. The silver voltage taps were placed at the 2nd turn and 3.5th turn. The distance between two taps is about 150 mm. The heat treatment was performed at 850 °C for 0.5 h in an Ar atmosphere. After the heat treatment, the SPCs were impregnated with epoxy resin. The transport properties of all SPCs were firstly tested at liquid helium and zero external magnetic field by the Institute of Plasma Physics, Chinese Academy of Sciences. Two weeks later, some of SPCs were inserted coaxially in the 38 mm bore of a 25 T resistive magnet at the High Magnetic Field Laboratory, Chinese Academy of Sciences. Figure 2 (b) and (c) show the Ba122 SPC before the testing and the 25 T resistive magnet

*3.2. Results and discussions*

Transport critical current $I_c$ of SPC at 4.2 K was measured by the standard four-probe method with a criterion of 1 $\mu V$/cm. Therefore, the critical criterion is 1 $\mu V$/cm × 15 cm=15 $\mu V$. At zero external magnetic field, the $I_c$ values are 76 and 77 A, when the excitation rates are 50 and 70 A/min, respectively. These $I_c$ values are two times larger than that of our previous DPC [9]. In order to evaluate the transport properties of Ba122 SPC at high field, the same coil was tested again at liquid helium and 25 T high field resistive magnet. The $I_c$ value is about 66 A at zero external magnetic field, which is slightly decreased. One main reason is maybe that the SPC was exposed to air for two weeks. Figure 3 shows the $I_c$ data of SPC with the magnetic field, including the self-field and the external magnetic field. In case that the self-field generated by the straight tape is small, the transport current of short Ba122 tape with the external magnetic field was also added in Fig. 3. When the inner diameter of this SPC is as small as 30 mm, its $I_c$ value is still as high as 35 A at 10 T, which is about half of that of the straight short sample. For our previous DPC [9], the $I_c$ value was reduced from 23 A (0 T) to 5 A (5 T) at 6 K, even the inner diameter of the DPC was as large as 71.5 mm. One possible reason is that a violent vibration happened to the tape during the insulator wrapping process, which is very harmful for the brittle superconducting core. These results indicate that the 7-filamentary Ba122/Ag/AgMn tape with 0.33 mm thickness was not damaged seriously during making coil, and that the non-insulation winding together with the stainless-steel tape is suitable to the iron-based superconductor.

From figure 3, we can also see that the $I_c$ of Ba122 SPC is also independent on magnetic field, like the short tape. However, though it is a promising improvement on our previous work [9], the discrepancy between the short sample data and the coil data is still considerable. There are some possible reasons for the transport current degradation of the SPC. 1) The strength of this Ba122/Ag/AgMn tape is still not strong enough for the coil winding process. It is helpless for the brittle superconducting core once a violent force happened to the tape. Fortunately, IBS wires or tapes can be fabricated by using stronger metals, such as Cu, Monel, stainless steel, etc. 2) There are maybe some low $I_c$ points in the long wire. It is needed to further test $I_c$ uniformity of the long wire before it can be used for coil winding. 3) The heat treatment of the short tape and the SPC was different. According to our experience, the better temperature is 880 °C rather than 850 °C, but the former is closer to the melting point of AgMn.

## 4. Conclusion

In conclusion, we successfully made 7-filamentary Ba122/Ag/AgMn single pancake coil with 30 mm inner diameter. This SPC was made using non-insulation Ba122/Ag/AgMn tape together with stainless steel tape by the wind-and-react process. The transport properties of Ba122 SPC was firstly tested at liquid helium and 24 T external magnetic field. Like the straight Ba122 short tape, the transport current of SPC was also independent on the background field and still large under high magnetic field. For example, the transport $I_c$ value of SPC was still 26 A at 24 T background field, which is about 40% of that at zero external magnetic field. These results suggest that the iron-based superconductors are very promising for high-field magnet applications.

**Acknowledgments**

The authors would like to thank Prof. S. Awaji from Sendai, and Miss Chao Tian, Dr. Fang Liu, Prof. Huajun Liu, Mr. Chuanyin Xi, Prof. Li Pi from Hefei, and Dr. Chao Yao, Mr. Yanchang Zhu, Mr. Zhe Cheng, Mr. Shifa Liu, Mr. Liu Li from IEE CAS, and Dr. Shaoqing Wei, Ms. Yingzhe Wang, Ms. Lingling Gong, Mr. Zhen Zhang, Prof. Yifang Wang from IHEP CAS for their great support of this work and helpful discussions. A portion of this work was performed on the Steady High Magnetic Field Facilities, High Magnetic Field Laboratory, CAS. This work is partially supported by the National Natural Science Foundation of China (Grant Nos. 51577182 and U1832213), the Beijing Municipal Science and Technology Commission (Grant No.

Z171100002017006), the Beijing Municipal Natural Science Foundation (Grant No. 3182039), the Key Research Program of Frontier Sciences，CAS (QYZDJ-SSW-JSC026), the Strategic Priority Research Program of Chinese Academy of Sciences (Grant No. XDB25000000).

**Table 1.** Specification of $Ba_{0.6}K_{0.4}Fe_2As_2$/Ag/AgMn tape

| Parameter | Unit | Value |
|---|---|---|
| Width | mm | 4.5 |
| Thickness | mm | 0.33 |
| Number of filament | | 7 |
| Non-SC/SC ratio | | 5.0 |
| $T_c^{onset}$ | K | 37.8 |
| $T_c^{zero}$ | K | 36.9 |
| $I_c$ @ 4.2 K, 10 T | A | 72 |

Table 2. Specification of Ba122 single pancake coil

| Parameter | Unit | Value |
|---|---|---|
| Inner diameter | mm | 30 |
| Outer diameter | mm | 34.8 |
| Height | mm | 4.62 |
| Thickness of stainless steel tape | mm | 0.1 |
| Turns |  | 4.5 |
| Total length of IBS wire | mm | 450 |

**Captions**

Figure 1. The magnetic field dependence of transport $J_c$ data at 4.2 K for 7-filamentary Ba122/Ag/AgMn tape. The inset shows the optical micrograph on transverse cross section of heat-treated 7-filamentary Ba122/Ag/AgMn tape.

Figure 2. The outer view of Ba122 single pancake coil with 30 mm inner diameter (a), the sintered Ba122 SPC (b) and the 25 T resistive magnet (c).

Figure 3. Magnetic field dependence of transport critical current at 4.2 K for Ba122 straight tape and SPC.

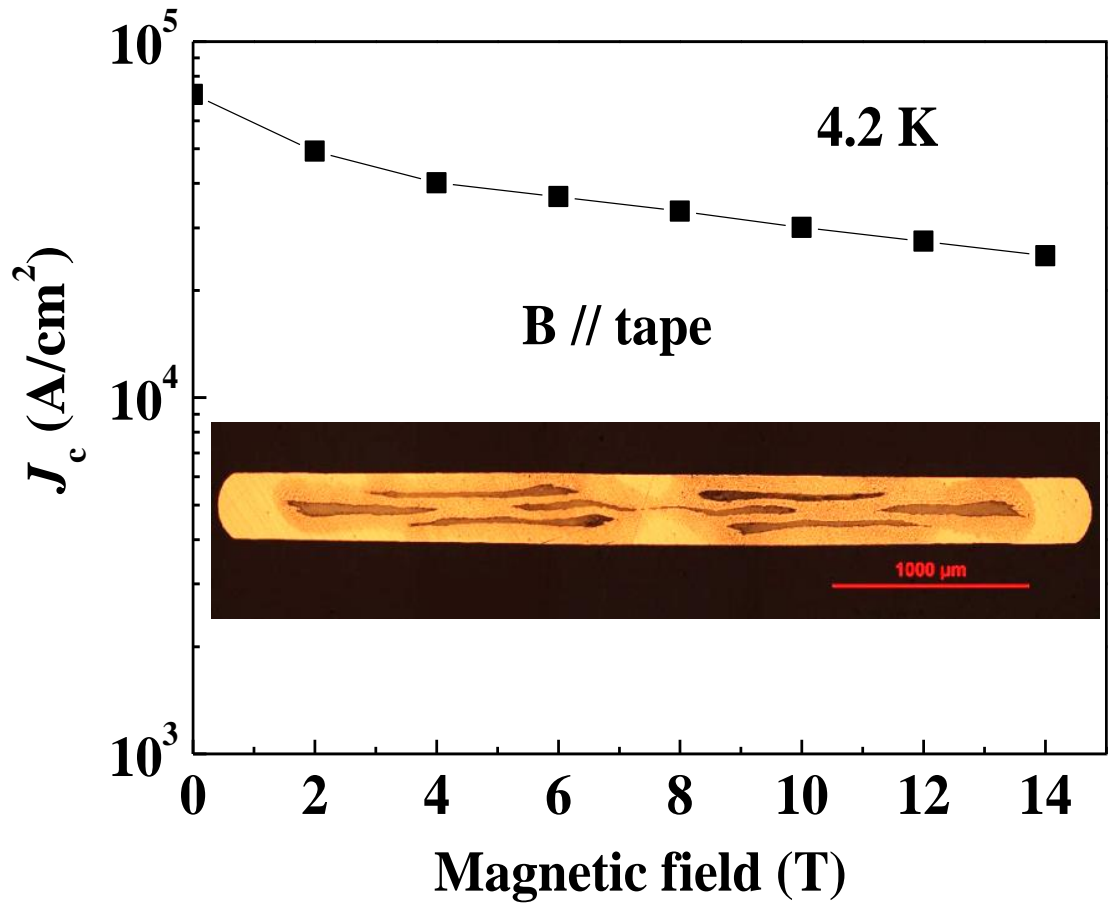

Figure 1. Wang et al

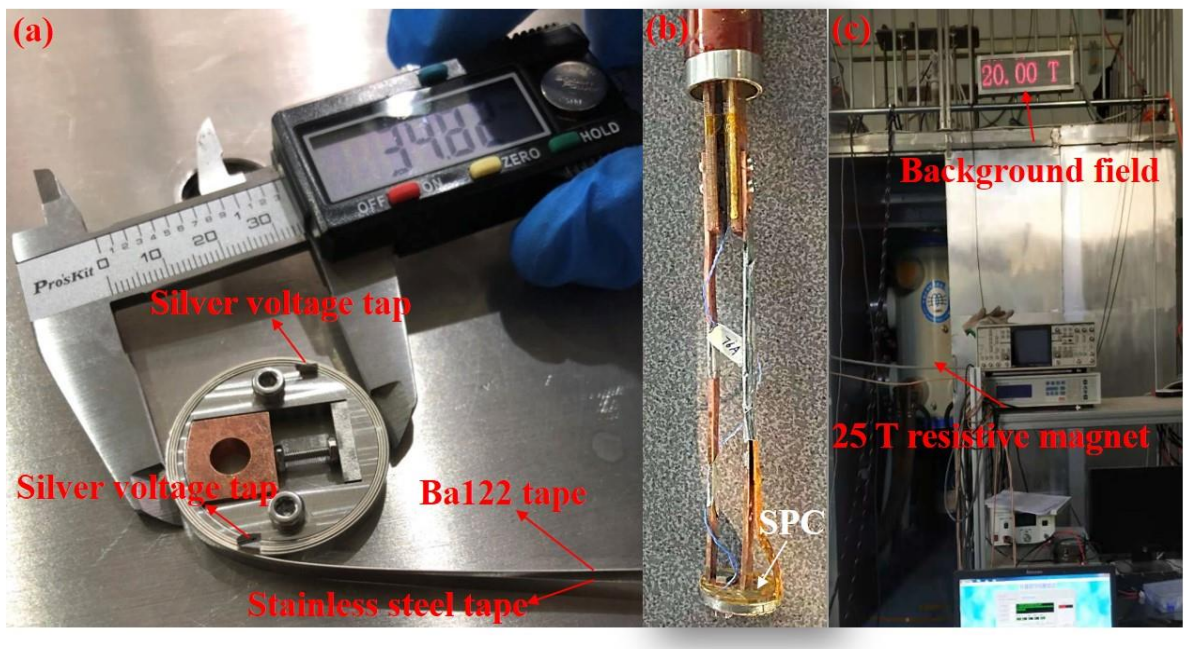

Figure 2. Wang et al

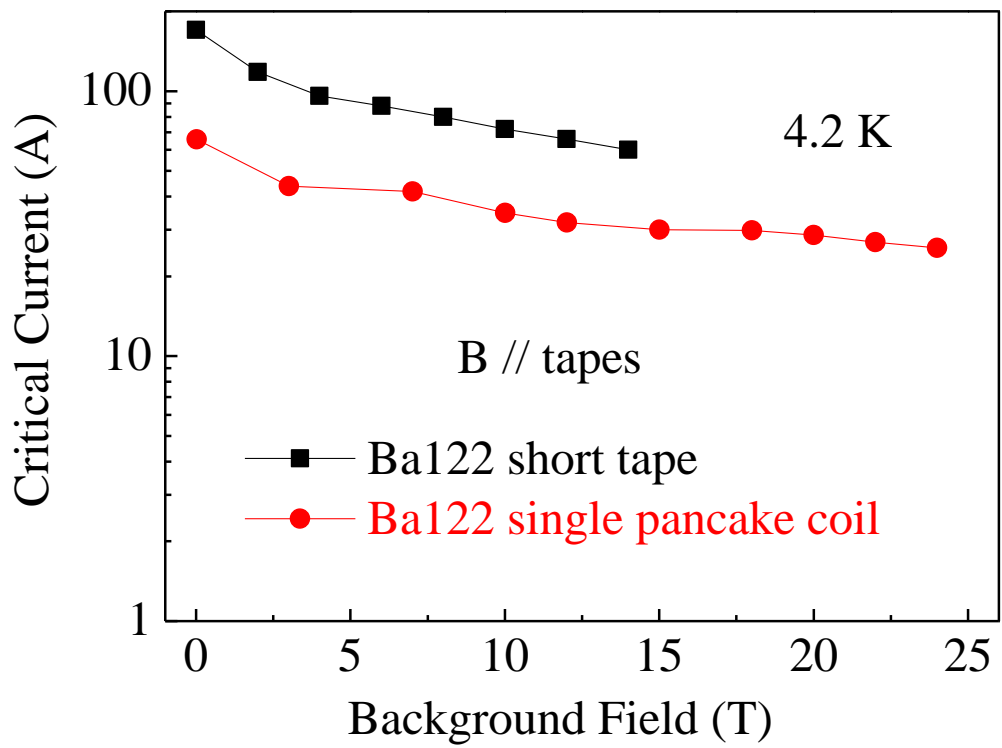

Figure 3. Wang et al